\documentclass[12pt]{iopart}

\usepackage[dvips]{graphicx}
\usepackage{lscape}

\newcommand{\bm}[1]{\mbox{\boldmath $#1$}}
\newcommand{\ds}{\displaystyle}
\newcommand{\st}{\scriptstyle}
\newcommand{\sz}{\scriptsize}

\begin{document}

\title{Optimal map of the modular structure of complex networks}

\author{A~Arenas$^{1,2}$\footnote{Author to whom any correspondence should be addressed},
J. Borge-Holthoefer$^1$, S~G\'omez$^1$ and G~Zamora-L\'opez$^3$}

\address{$^1$ Departament d'Enginyeria Inform\`{a}tica i Matem\`{a}tiques,
Universitat Rovira i Virgili,
43007 Tarragona, Spain}

\address{$^2$ Lawrence Berkeley National Laboratory, Berkeley, CA 94720, USA}

\address{$^3$ Interdisciplinary Center for Dynamics of Complex System, University of Potsdam, 14415 Potsdam, Germany}

\eads{\mailto{alexandre.arenas@urv.cat},
\mailto{javier.borge@urv.cat}
\mailto{sergio.gomez@urv.cat} and \mailto{gorka\_agnld@yahoo.es}}

\begin{abstract}
Modular structure is pervasive in many complex networks of interactions observed in natural, social and technological sciences. Its study sheds light on the relation between the structure and function of complex systems. Generally speaking, modules are islands of highly connected nodes separated by a relatively small number of links. Every module can have contributions of links from any node in the network. The challenge is to disentangle these contributions to understand how the modular structure is built. The main problem is that the analysis of a certain partition into modules involves, in  principle, as many data as number of modules times number of nodes. To confront this challenge, here we first define the contribution matrix, the mathematical object containing all the information about the partition of interest, and after, we use a Truncated Singular Value Decomposition to extract the best representation of this matrix in a plane. The analysis of this projection allow us to scrutinize the skeleton of the modular structure, revealing the structure of individual modules and their interrelations.
\end{abstract}

\maketitle

\section{Introduction}

The concept of modular structure in real complex networks \cite{gn} is revolutionizing the understanding of the evolution of complex systems \cite{vespi}. Many efforts have been devoted to its automatic detection \cite{ng,palla,ddda}, however very little is known yet about the actual skeleton of the detected modules that build the network. This skeleton promises to be relevant to understand why physical processes in complex networks, such as synchronization \cite{physrep}, present emergent phenomena that are affected by the existence of topological barriers between modules. We still miss fundamental tools to anticipate these phenomena from a topological perspective.
The current work is intended to provide network scientists with novel tools to screen the modular structure. The comprehension of modular structure in networks necessarily demands the analysis of the contribution of each one of its constituents (nodes) to the modules. Recently, Guimer{\`a} et al. \cite{rogernat,rogerair} advanced on this issue proposing two descriptors to characterize the modular structure: the $z$-score (a measure of the number of standard deviations a data point is from the mean of a data set) of the internal degree of each node in its module, and the participation coefficient ($P$) defined as how the node is positioned in its own module and with respect to other modules. Given a certain partition, the plot of nodes in the $z$--$P$ plane admits an heuristic tagging of nodes' role. The success of this representation relies on a consistent interpretation of topological roles of nodes according to the specific data analyzed.

Here we introduce a formalism to reveal the characteristics of networks at the {\em topological mesocale}, where the representation of the network is viewed as a set of interconnected modules. We propose a method, based on linear projection theory, to study the modular structure in networks that enables a systematic analysis and elucidation of its skeleton. First, we construct a matrix containing all the information about the modular structure, and  second, we find an optimal dimensional reduction of the information contained in it. In particular, we present the optimal mapping of the information of the modular structure (in the sense of least squares) in a two-dimensional space. The method has been applied to synthetic and real networks. The statistical analysis of the geometrical projections allow to characterize the structure of individual modules and their interrelations in a unified framework.

The paper is structured as follows. In section 2, we present the motivation of the method and the main findings to interpret the outcome. In section 3, the method is illustrated with synthetic networks whose structure is controlled. Finally, in section 4, the method is tested in real networks and an explanation of the results is offered.

\section{Projection of the modular structure}

A complex network (weighted or unweighted, directed or undirected) can be represented by its graph matrix $\bm{W}$, whose elements $W_{ij}$ are the weights of the connections from any node $i$ to any node $j$. Assuming that a certain partition of the network into modules is available, we plan to analyze this coarse grained structure. Note that the partition can be obtained by any method, and that the method we propose based on modularity \cite{ng} is a possibility. The main object of our analysis is the \emph{Contribution matrix} $\bm{C}$, of $N$ nodes to $M$ modules. The rows of $\bm{C}$ correspond to nodes, and the columns to modules. The analysis of this matrix is the focus of our research. The elements $C_{i\alpha}$ are the number of links that node $i$ dedicates to module $\alpha$, and can be easily obtained as the matrix multiplication between $W_{ij}$ and the {\em partition matrix} $\bm{S}$:
\begin{equation}
C_{i\alpha} = \sum_{j=1}^N W_{ij} S_{j\alpha}
\end{equation}
where $S_{j\alpha} = 1$ if node $j$ belongs to module $\alpha$, and $S_{j\alpha} = 0$ otherwise. The goal is to reveal the structure of individual modules, and their interrelations, from the matrix $\bm{C}$. To this end, we propose to deal with the high dimensionality of the original data by constructing a two-dimensional map of the contribution matrix, minimizing the loss of information in the dimensional reduction, and making it more amenable to further investigation.

\subsection{Singular Value Decomposition of the modular structure}

The approach developed here consists in the analysis of $\bm{C}$ using Singular Value Decomposition \cite{svd} (SVD). It stands for the factorization of a rectangular $N$-by-$M$ real (or complex) matrix as follows:
\begin{equation}
  \bm{C}=\bm{U} \bm{\Sigma} \bm{V}^{\dag}
\end{equation}
where $\bm{U}$ is an unitary $N$-by-$N$ matrix, $\bm{\Sigma}$ is a diagonal $N$-by-$M$ matrix and $\bm{V}^\dag$ denotes the conjugate transpose of $\bm{V}$, an $M$-by-$M$ unitary matrix. This decomposition corresponds to a rotation or reflection around the origin, a non-uniform scale represented by the {\em singular values} (diagonal elements of $\bm{\Sigma}$) and (possibly) change in the number of dimensions, and finally again a rotation or reflection around the origin. This approach and its variants have been extraordinarily successful in many applications \cite{svd}, in particular for the analysis of relationships between a set of documents and the words they contain. In this case, the decomposition yields information between word-word, word-document, and document-document  semantic associations, the technique is known as Latent Semantic Indexing \cite{berry}, and Latent Semantic Analysis \cite{landauer}. Our scenario is quite similar to this, where nodes resemble words, and modules resemble documents. We devise that a similar approach will help to unravel the relations between nodes' contributions and modules of a certain partition.

\subsection{An optimal 2D map of the modular structure of networks}

A practical use of SVD is dimensional reduction approximation, also known as Truncated Singular Value Descomposition (TSVD). It consists in keeping only some of the largest singular values to produce a least squares optimal, lower rank order approximation (see Appendix). In the following we will consider the best approximation of $\bm{C}$ by a matrix of rank $r=2$.

The main idea is to compute the projection of the contribution of nodes to a certain partition (rows of $\bm{C}$, namely $\bm{n}_i$ for the $i$-th node) into the space spanned by the first two left singular vectors,
 the projection space $\mathcal{U}_2$ (see Appendix). We denote the projected contribution of the $i$-th node as $\bm{\tilde{n}}_i$. Given that the transformation is information preserving \cite{chu05}, the map obtained gives an accurate representation of the main characteristics of the original data, visualizable and, in principle, easier to scrutinize. Note that the approach we propose has essential differences with classical pattern recognition techniques based on TSVD such as Principal Components Analysis (PCA) or, equivalently, Karhunen-Loeve expansions. Our data (columns of $\bm{C}$) can not be independently shifted to mean zero without loosing its original meaning, this restriction prevents the straightforward application of the mentioned techniques, and also differentiates our work from the modern techniques for the analysis of gene expression patterns \cite{genes1,genes2}.

\begin{figure}[!tpb]
  \begin{center}
\includegraphics*[angle=0,width=.80\textwidth]{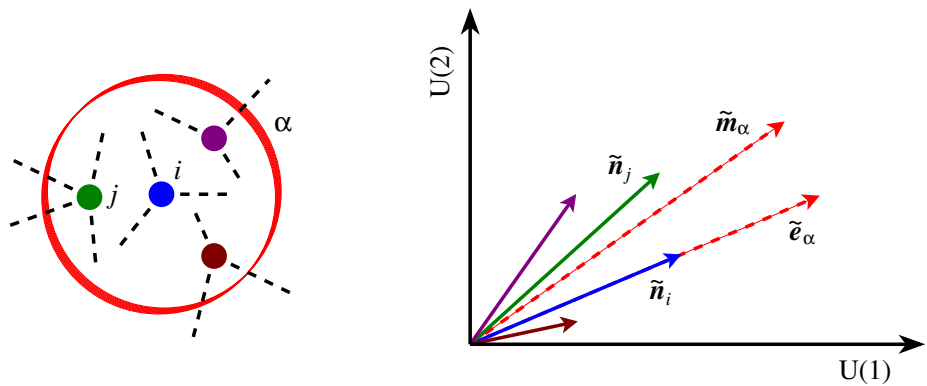}
  \end{center}
\caption{Geometrical scheme of the TSVD. The intramodular projection of module $\alpha$, $ \bm{\tilde{e}}_{\alpha}$ is the direction where all internal nodes lay (in the plot node $i$). The node contribution projections $\bm{\tilde{n}}$ are represented by vectors in different colors. Finally, the modular projection $\bm{\tilde{m}}_{\alpha}$ is computed as the vector sum of all the node contribution projections belonging to it. Note that the intramodular projection and the modular projection do not coincide, the differences between both inform about the cohesiveness of the module.}
\label{f1}
\end{figure}

The main problem when using SVD relies always on the interpretation of its outcome. The combination of data in the process makes difficult a direct comparison between input and output. To overcome this problem, we point out the following geometrical properties of the projection of the rows of $\bm{C}$ we have defined (see Appendix for a mathematical description):

\begin{enumerate}
\item Every module $\alpha$ has an intrinsic direction $\bm{\tilde{e}}_{\alpha}$ in the projection space $\mathcal{U}_2$ corresponding to the line of the projection of its internal nodes (those that have links exclusively inside the module). We call these directions {\em intramodular projections}. This property is essential to discern among modules that are cohesive, in the sense that the majority of nodes project in this direction, from those modules which are not.

\item Every module $\alpha$ has a distinguished direction $\bm{\tilde{m}}_{\alpha}$ in the projection space $\mathcal{U}_2$ corresponding to the vector sum of the contributions of all its nodes. We call these directions {\em modular projections}. The modular projection is relevant when compared to the intramodular projection because their deviations inform about the tendency to connect with other modules. Note that $e_\alpha$ and $m_\alpha$ are equal only if the module is disconnected from the rest of the network.

\item Any node contribution projection $\bm{\tilde{n}}_i$ is a linear combination of intramodular projections, being the coefficient of each one proportional to the original contribution $C_{i\alpha}$ of links of the node $i$ to each module $\alpha$. This property comes from the linearity of the projection, and expresses the contribution of nodes to the modules to which they are connected to.

\end{enumerate}
Consequently, from (i) and (iii), we can classify nodes. Nodes with only internal links have a distance to the origin proportional to its degree (or strength). Nodes with internal and external links, separate from the intramodular projection proportionally to their contributions to other modules. From (ii) we can classify modules. Modules that have close modular projections are more interrelated.
These geometrical facts are the key to relate the outcome of TSVD and the original data in our problem, see Fig.~\ref{f1}.

\subsection{Structure of individual modules}

To study the structure of individual modules we concentrate on the analysis of the projection of nodes' contributions in the plane $\mathcal{U}_2$. Keeping in mind the geometrical properties (i) and (iii) exposed above, we propose to extract structural information relative to each module by comparing the map of nodes' contributions to the intramodular projection directions. To this end it is convenient to change to polar coordinates, where for each node $i$ the radius $R_i$ measures
the length of its contribution projection vector $\bm{\tilde{n}}_i$,
and $\theta_i$ the angle between $\bm{\tilde{n}}_i$ and the horizontal axis.
We also define $\phi_i$ as the absolute distance in angle between $\bm{\tilde{n}}_i$ and the intramodular projection $\bm{\tilde{e}}_{\alpha}$ corresponding to its module $\alpha$, i.e. $\phi_i=|\theta_i - \theta_{\bm{\st\tilde{e}}_{\alpha}}|$.

\begin{figure}[!tpb]
  \begin{center}
\includegraphics*[angle=0,width=.40\textwidth]{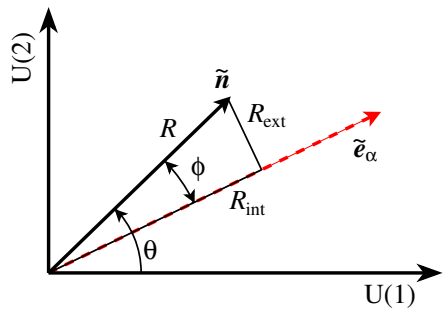}
  \end{center}
\caption{Schematic plot of the coordinates proposed to study the structure of individual modules.  The relative distance of a node from its module is captured by the angle $\phi$. The respective components
$R_{\mbox{\sz int}}$ and $R_{\mbox{\sz ext}}$ are depicted.}
\label{f2}
\end{figure}

Using these coordinates $R$--$\phi$ we find a way to interpret correctly the map of the contribution matrix in  $\mathcal{U}_2$: i) $R_{\mbox{\sz int}}=R\cos\phi$ informs about the internal contribution of nodes to its corresponding module, as well as to the contribution to its own module by connecting to others. To clarify the latter assertion, let us assume a node $i$ belonging to a module $\beta$ has connections with the rest of modules in the network. Given that this connectivity pattern is a linear combination of intramodular directions $ \bm{\tilde{e}}_{\alpha}$ , the vector sum implies that connecting with modules $\alpha$ having $|\theta_{\bm{\st\tilde{e}}_{\beta}}- \theta_{\bm{\st\tilde{e}}_{\alpha}}| > \pi/2$ decreases the module $R$, and vice versa. ii) $R_{\mbox{\sz ext}}=R \sin\phi$ informs about the deviation (as the orthogonal distance) of each node to the contribution to its own module, see Fig.~\ref{f2}. It is also possible to study the spreading of $\phi$ by using other descriptors proposed in the context of synchronization \cite{rosenblum}.

We explore the internal structure of modules using the values of $R_{\mbox{\sz int}}$, and the boundary structure of modules using $R_{\mbox{\sz ext}}$. Using descriptive statistics one can reveal and compare the structure of individual modules. Provided that the distribution of contributions is not necessarily Gaussian, an exploration in terms of z-scores is not convenient. Instead we use box-and-whisker charts for the variables, depicting the principal quartiles and the outliers (defined as having a value more than 1.5 IQR lower than the first quartile or 1.5 IQR higher than the third quartile, where IQR is the Inter-Quartile Range).

The boxplots for the data of each module in the variable
$R_{\mbox{\sz int}}$ allow for a visualization of the heterogeneity in the contribution of nodes building their corresponding modules, and an objective determination of distinguished nodes on its structure (outliers). Consequently, the boxplots in
$R_{\mbox{\sz ext}}$
inform about the heterogeneity in the boundary connectivity. Nodes with links in only one module are not considered in this statistics because they do not provide relevant information about the boundaries (they have $\phi = 0$), only nodes that act as bridges between modules are taken into account. Considering internal nodes in this statistics would eventually produce a collapse of the quartiles to zero. Assuming that every module devotes some external links (otherwise they would be disconneted), the width of the boxes in this plot is proportional to the heterogeneity of such efforts. If only one node makes external connections, then the boxplot has zero width. Moreover, given two boxes equally wide, their position (median) determines which module contributes more to keeping the whole network connected.

\subsection{Interrelations between modules}
\begin{figure}[!tpb]
  \begin{center}
\includegraphics*[angle=0,width=.75\textwidth]{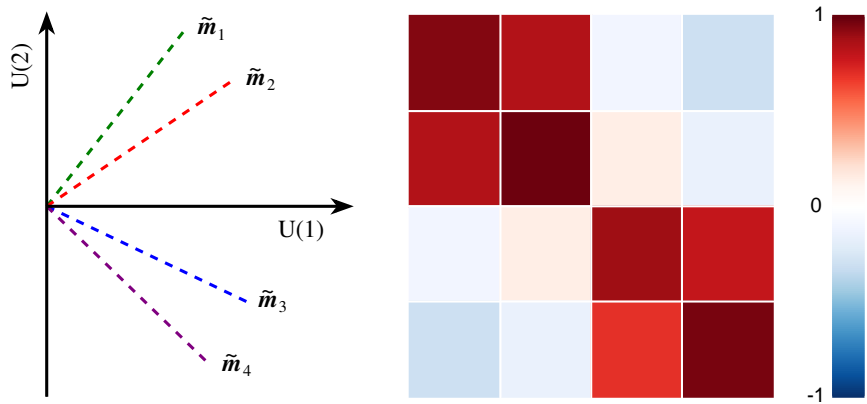}
  \end{center}
\caption{Schematic plot of the interrelation between the modular projections of 4 modules. The matrix represents the overlap computed as the scalar product between directions.}
\label{f3}
\end{figure}
The analysis of the interrelations between modules is performed at the coarse grained level of its modular projections. The modular projections  $\bm{\tilde{m}}_{\alpha}$ are aggregated measures of the nodes' contribution to their particular module. The normalized scalar product of modular projections provide a measure of the interrelations (overlapping) between different modules. A representation of these data in form of a matrix ordered by the values of $\theta_{\bm{\st\tilde{m}}_{\alpha}}$ reveals the actual skeleton of the network at the topological mesoscale, see Fig.~\ref{f3}.

\section{Application to synthetic networks}

\begin{figure}[!tpb]
  \begin{center}
  \begin{tabular}{ll}
    {\bf a)} \hspace{25pt} Cliques line partition 2 modules
    \\
    \mbox{\includegraphics*[width=.75\textwidth]{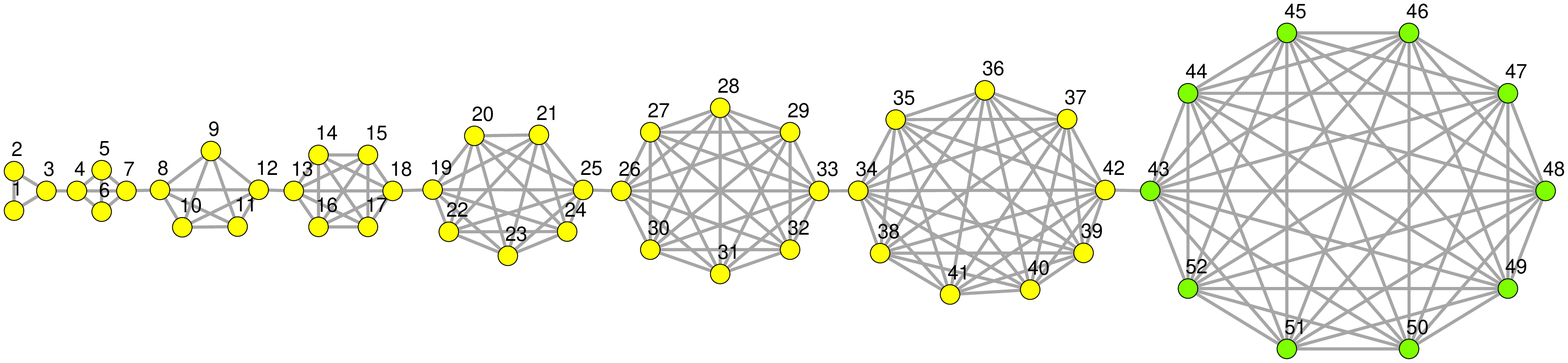}}
    \\
     {\bf b)} \hspace{25pt} Cliques line partition 8 modules
    \\
    \mbox{\includegraphics*[width=.75\textwidth]{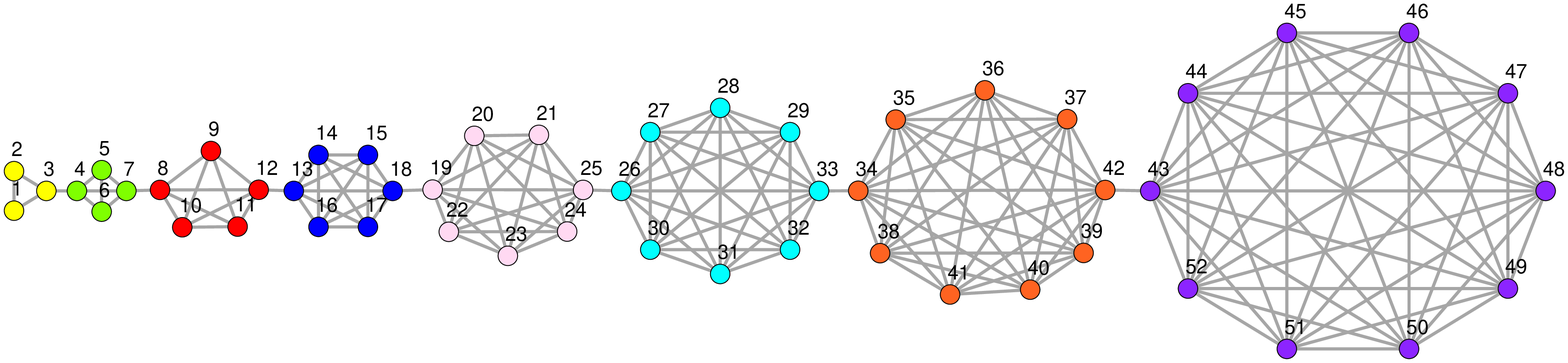}}
    \\

    {\bf c)} \hspace{25pt} 2-modules
    \\
    \multicolumn{2}{l}{\mbox{\includegraphics*[width=.75\textwidth]{fig_4c.eps}}}
    \\ \\
    {\bf d)} \hspace{25pt} 8-modules
    \\
    \multicolumn{2}{l}{\mbox{\includegraphics*[width=.75\textwidth]{fig_4d.eps}}}
  \end{tabular}
  \end{center}
\caption{Optimal map of the modular structure for the optimal partition of the cliques network partitioned in two modules ({\bf a}) and the cliques network partitioned in eight modules ({\bf b}), each color corresponds to a different module of the given partition. In ({\bf c}) and ({\bf d}) we plot the projected space spanned by the two left singular vectors of the TSVD, $\mathcal{U}_2$ (left), and its transformation to polar coordinates $R$--$\theta$ (right), for each network. Dashed lines mark the directions of intramodular projections of each module. In d) right we present a zoom in $\theta$ for better visual inspection.
}
\label{f4}
\end{figure}

We start applying the methodology of analysis to synthetic networks, having control of the whole network structure. First, we analyze a network built up from cliques of different sizes, we consider a line of cliques from size 3 to 10, joined only by a unique link between them. We will consider two different partitions to test the method. The first partition consists of a module containing the larger clique, and another containing the rest of the cliques, see Fig.~\ref{f4}a. In the second partition each clique forms a module, see Fig~\ref{f4}b. The plots Fig.~\ref{f4}c,d (left) show the projections of the nodes' contributions in the plane spanned by the two first right singular vectors $\mathcal{U}_2$, as well as the intramodular projections of each module in this plane. The data in $\mathcal{U}_2$ are transformed to polar coordinates for a better visualization and simpler analysis, see Fig.~\ref{f4}c,d (right). The structure of these plots will be repeated in the next examples.

Projecting the contribution matrix corresponding to the partition in two modules Fig.~\ref{f4}c, we observe clearly the relation in connectivity between nodes and the structure of both modules. The two distinguished nodes that connect both modules lay out of the intramodular projections, while the rest of nodes lay exactly on this direction. The different positions within the intramodular projections correspond to the degree of each node, nodes with identical contribution project to the same position. For the second partition, Fig~\ref{f4}d, the modules of size 3 to 9, are concentrated around a similar direction while the clique of size 10 is separated from the rest. In the plot we have zoomed the regions in the $R$-$\theta$ around the directions where nodes project. For every module the projection reflects two positions: one exactly on the intramodular direction corresponding to the internal nodes of the clique and another corresponding to the node that acts as a connector with the following clique. The connectors towards the precedent clique (of lower size) are indistinguishable at the resolution of the plot, but also lay in a different direction.

Following the test, now we apply the method to a model of network with a well defined community structure that
has been used as a benchmark for different community detection
algorithms \cite{ddda}, proposed by Girvan and Newman
\cite{ng}. In that model the authors construct a network
of 128 nodes as a set of 4 communities, each one formed by  32
nodes. Fixing the mean number of links per node at a value of 16,
the parameter describing the sharpness of the community
distribution is $z_{\mbox{\scriptsize in}}$, the average number of links within the
community. A generalization of this model was proposed in \cite{physicad} to
include several
hierarchical levels of communities.
\begin{figure}[!tpb]
  \begin{center}
  \begin{tabular}{l}
    {\bf a)} \hspace{25pt}
    \\
    \hspace{110pt} \mbox{\includegraphics*[angle=90,width=.30\textwidth,height=.20\textwidth]{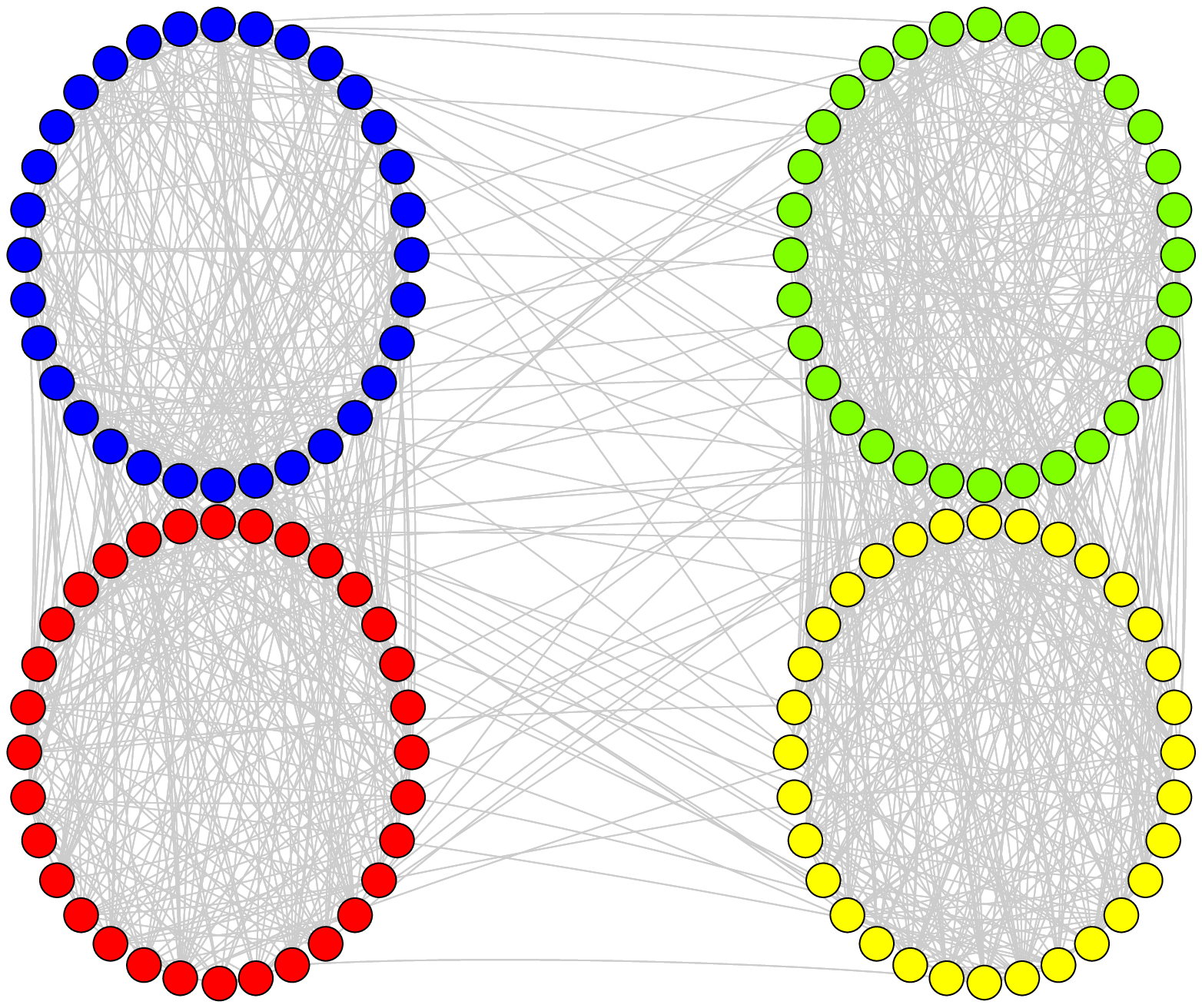}}
    \\ \\
    {\bf b)} \hspace{25pt}
    \\
    \mbox{\includegraphics*[angle=0,width=.75\textwidth]{fig_5b.eps}}
  \end{tabular}
  \end{center}
\caption{Analysis of a random homogeneous hierarchical network with community structure, see text for details. a) Network structure.
b) Projection as explained in Fig.~\ref{f4}.}
\label{f5}
\end{figure}
The hierarchy is defined as follows:  we take a set of $N$ nodes
and divide it into $n_1$ groups of equal size; each of these
groups is then divided into $n_2$ groups and so on up to a number
of steps $k$ which defines the number of hierarchical levels of
the network. Then we add links to the networks in such a way that
at each node we assign at random a number of $z_{1}$ neighbours
within its group at the first level, $z_{2}$ neighbours within
the group at the second level and so on. There remains the number
of links that each node has to the rest of the network;
that we will call $z_{\mbox{\scriptsize out}}$.  We construct a network with $N=128$ nodes, two hierarchical levels
with $n_1=2$, $n_2=2$, $z_1=5$, $z_2=10$ and $z_{\mbox{\scriptsize out}}=1$. Again the method resolves the modular structure
and individual contributions in the correct way, see Fig.~\ref{f5}. In Appendix D we also test the sensitivity and robustness of the method to
slight changes in the predefined partition.


\section{Application to real networks}

\begin{figure}[!tpb]
  \begin{center}
  \begin{tabular}{l}
    {\bf a)} \hspace{25pt}
    \\
     \hspace{70pt} \mbox{\includegraphics*[angle=0,width=.45\textwidth]{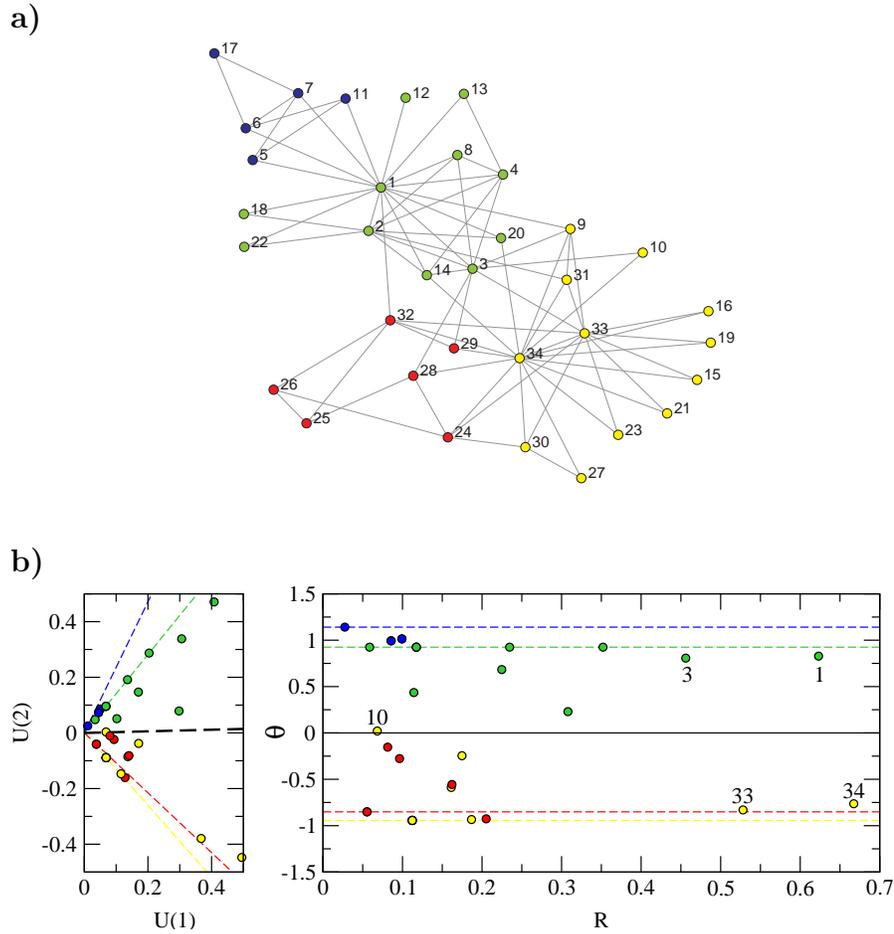}}
    \\ \\
    {\bf b)} \hspace{25pt}
    \\
    \mbox{\includegraphics*[angle=0,width=.75\textwidth]{fig_6b.eps}}
  \end{tabular}
  \end{center}
\caption{Analysis of the Zachary network for the four modules found by maximizing modularity. a) Network with each module represented in a different color. b) Projection as explained in Fig.~\ref{f4}.}
\label{f6}
\end{figure}

The first network analyzed is the Zachary's karate club network \cite{zachary} accounting for the study over two years of the friendships between 34 members of a karate club at a US university in 1970. The network in question was divided, at the end of the study period, in two groups after a dispute between the club's administrator (node 1) and the club's instructor (node 34), which ultimately resulted in the instructor leaving and starting a new club, taking about half of the original club's members with him. The partition we have used in our study corresponds to four modules resulting from optimizing modularity \cite{ng} using Extremal Optimization \cite{EO} and refined with Tabu search \cite{meso}, providing a value of modularity $Q=0.420$. After the projection,  see Fig.~\ref{f6}, we observe, nodes 1, 3 in the green module and 33, 34 in the blue module clearly distinguished by its value of $R$, denoting their important role in supporting the structure of both modules, however they are not the nodes that connect with other modules. It is also remarkable that node 10 lays half way of the modular directions of the larger modules assessing its unclassifiable nature (this node has been persistently misclassified by most of the community detection algorithms).

The proposed mapping is also applied to two other real networks, the worldwide air transportation network, and the  AS--P2P  Internet network. The airports network data set is composed of passenger flights operating in the time period November 1, 2000, to October 31, 2001 compiled by OAG Worldwide (Downers Grove, IL) and analyzed previously by Prof.\ Amaral's group \cite{rogerair}. It consists of 3618 nodes (airports) and 14142 links, we used the weighted network in our analysis. Airports corresponding to a metropolitan area have been collapsed into one node in the original database. The  AS--P2P  Internet data set considered is composed of autonomous systems (AS) \cite{caida} in the peer to peer (P2P) category, where two ASs freely exchange traffic between themselves and their customers, but do not exchange traffic from or to their providers or other peers \cite{caida2}. We complemented this data set with the geographic localization of the ASs, resulting in 1217 nodes and 4058 links. We have optimized modularity \cite{ng} to find good partitions of the networks in modules. We have used the partition corresponding to 26 modules and modularity $Q= 0.649$ for the airports network, and 12 modules and $Q= 0.387$ for the AS--P2P network. Note that any partition, not necessarily the one corresponding to optimal modularity, can be analyzed as described.

The interesting aspect of applying the analysis to these two data sets is twofold: first, since both are geo-referenced, it is possible to assign a tag to each module corresponding to geographic areas, and second, the modular structure of both networks is substantially  different, while the airports network evolution has been mainly shaped by two well defined continental blocks (USA and W Europe)\footnote{We denote N-S-E-W for the four cardinal points North, South, East and West respectively.}, the AS--P2P network has been built in a more homogeneous way. It is very interesting to observe how the AS-P2P network, following a sort of ``wiring optimization'', presents a community structure evenly distributed in areas covering a worldwide belt.

\begin{figure}[!tpb]
  \begin{center}
  \begin{tabular}{ll}
    {\bf a)} \hspace{25pt} Airports & \hspace{12pt} {\bf b)} \hspace{25pt} AS--P2P
    \\
    \mbox{\includegraphics*[width=.45\textwidth]{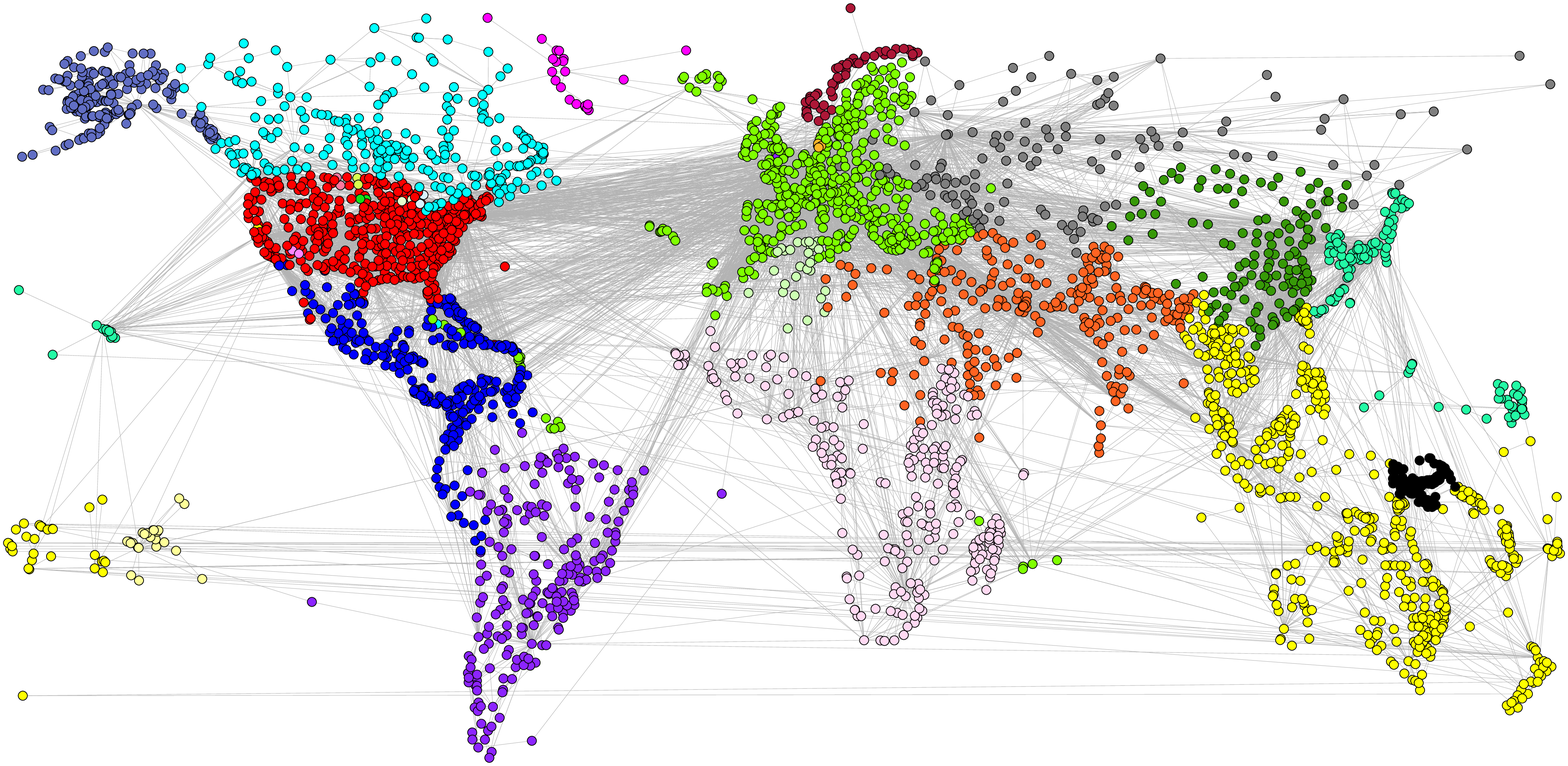}}
    &
    \begin{tabular}[b]{l}
    \mbox{\includegraphics*[width=.45\textwidth]{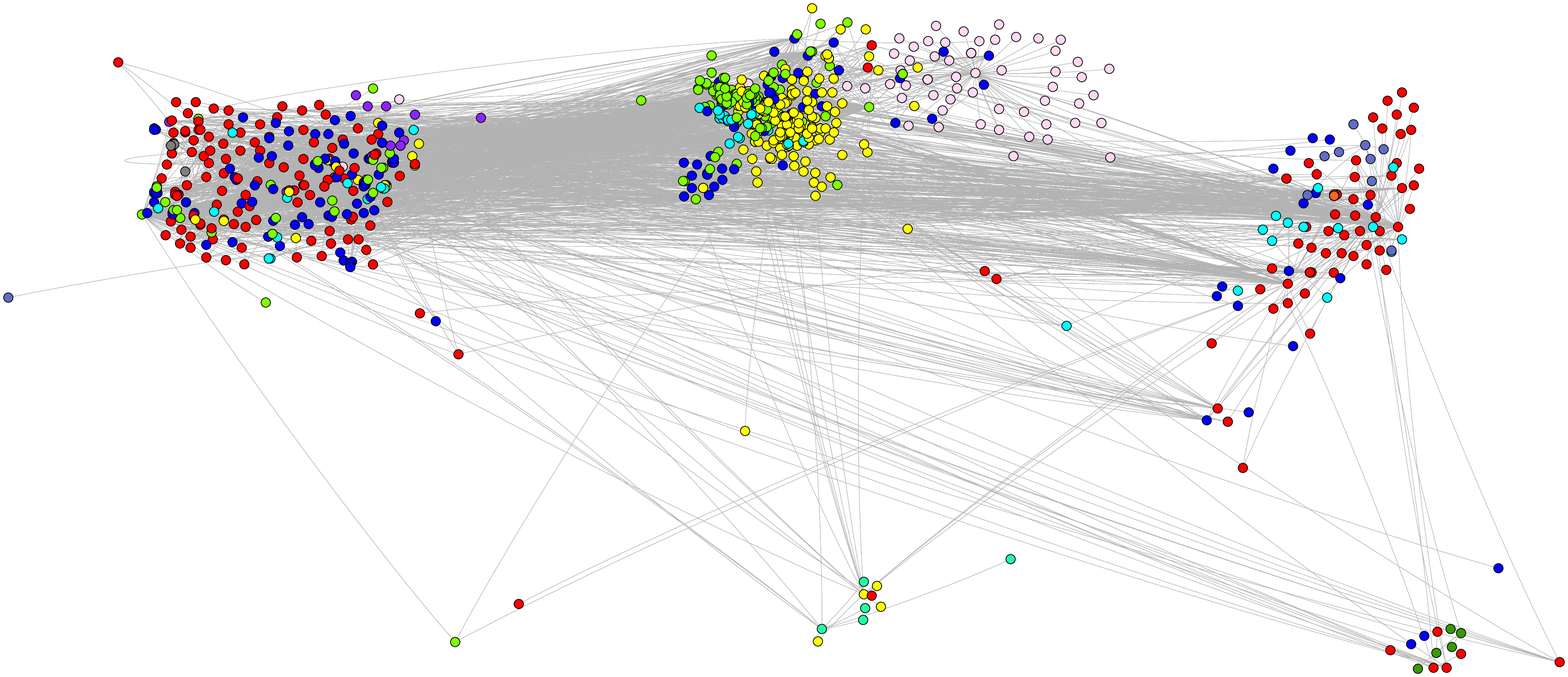}}
    \\
    \mbox{\rule{0pt}{5pt}}
    \end{tabular}
    \\ \\
    {\bf c)} \hspace{25pt} Airports
    \\
    \multicolumn{2}{l}{\mbox{\includegraphics*[width=.95\textwidth]{fig_7c.eps}}}
    \\ \\
    {\bf d)} \hspace{25pt} AS--P2P
    \\
    \multicolumn{2}{l}{\mbox{\includegraphics*[width=.95\textwidth]{fig_7d.eps}}}
  \end{tabular}
  \end{center}
\caption{Optimal map of the modular structure for the optimal partition of the airports network ({\bf a}) and the AS--P2P network ({\bf b}), each color corresponds to a different module of the given partition. In ({\bf c}) and ({\bf d}) we plot the projected space spanned by the two left singular vectors of the TSVD, $\mathcal{U}_2$ (left), and its transformation to polar coordinates $R$--$\theta$ (right), for each network. Dashed lines mark the directions of intramodular projections of each module. Nodes whose contribution is totally internal to a module project exactly on its corresponding dashed line. In the $R$--$\theta$ plot we have labelled certain distinguished nodes that also correspond to very important airports and ASs in the world. For the airports network we have magnified the area over $10^{-1}$ to identify the more important nodes in $R$. The loss of information associated to the two-dimensional projection is 18.2\% for the airports network and 15.8\% for the AS--P2P network.
}
\label{f7}
\end{figure}

In Fig.~\ref{f7}a,b, we plot the structure of the networks partitioned in modules, these conform the original data that compose our contribution matrices. The geographical location has been added to the plot for visualization purposes but it has not been used in the analysis. The plots Fig.~\ref{f7}c,d (left) show the projections of the nodes' contributions following the same structure of the precedent plots. The differences between both modular structures has clearly emerged in this projection, the airports network is basically polarized in two geographical areas, whereas in the AS--P2P network this polarization does not exist. We also see how different airports and ASs excel in their values of $R$ largely over the rest. This effect can be further developed by studying the structure of modules and their interrelations in each case.
\begin{landscape}
\begin{figure}[!tpb]
  \begin{center}
  \begin{tabular}{ll}
    {\bf a)} \hspace{19pt} Airports
    &
    {\bf b)} \hspace{19pt} AS--P2P
    \\ \\
    \mbox{\includegraphics*[width=.7\textwidth]{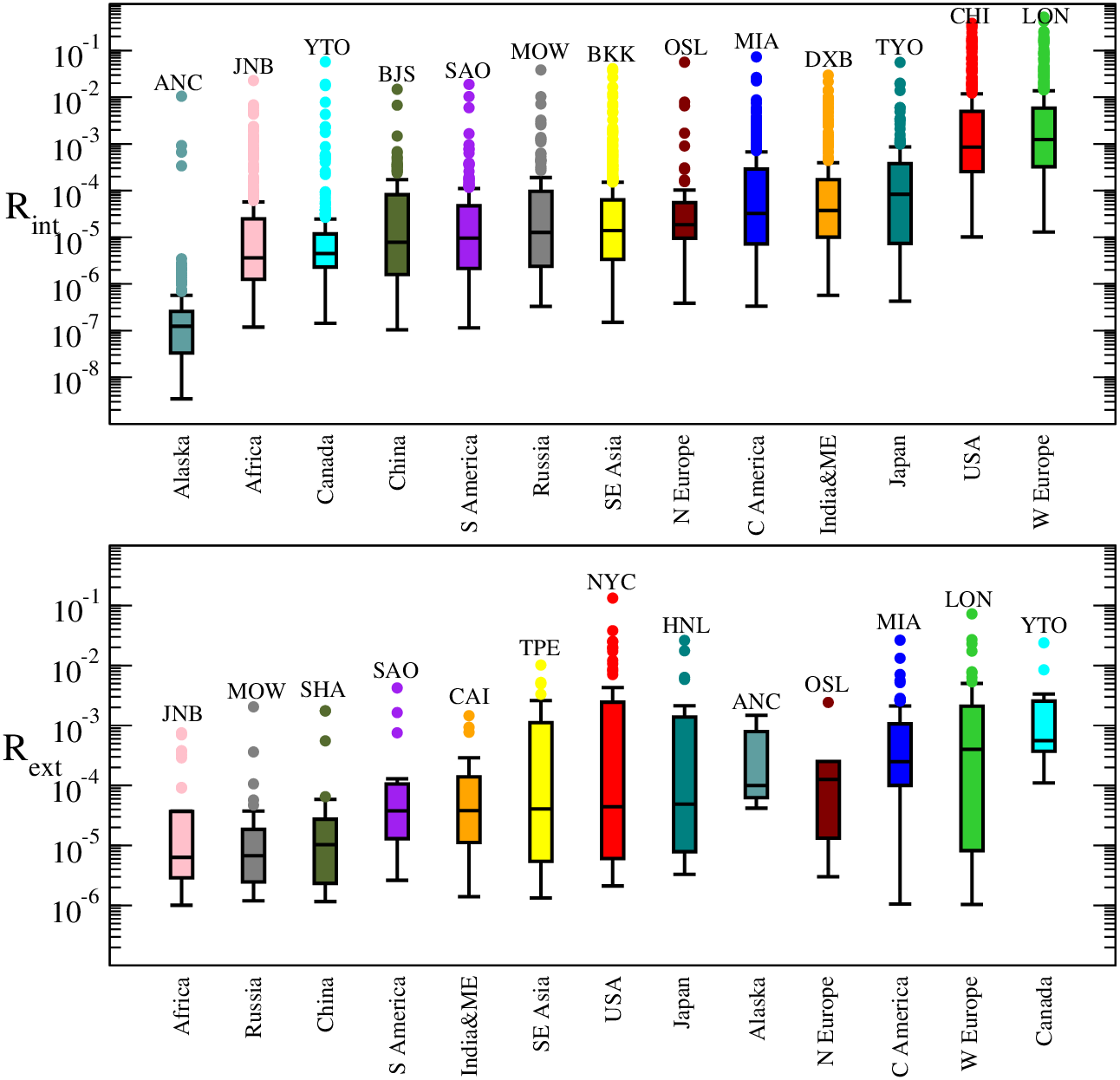}}
    &
    \mbox{\includegraphics*[width=.7\textwidth]{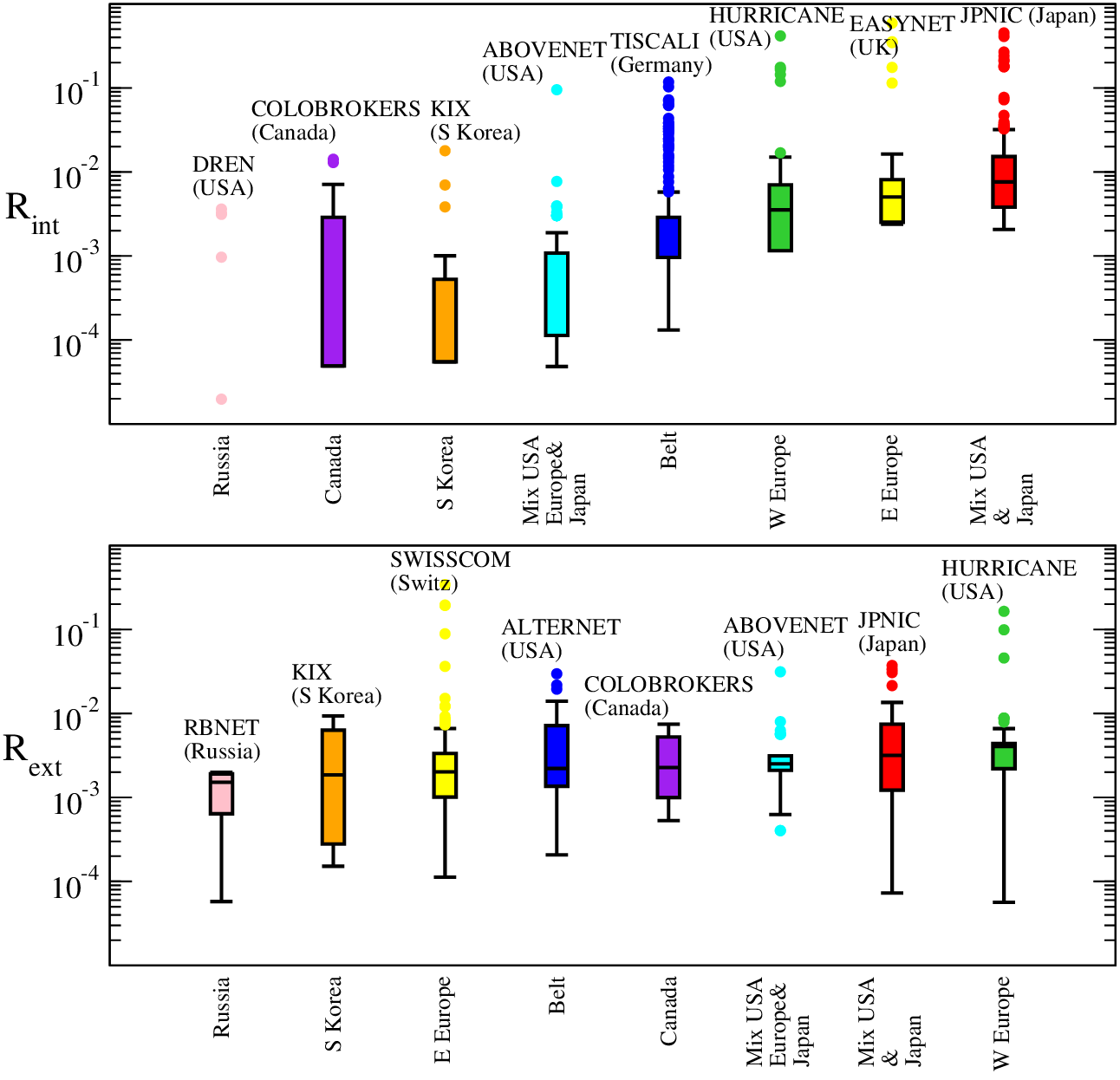}}
  \end{tabular}
  \end{center}
\caption{Box-and-whisker plots of $R_{\mbox{\sz int}}$ and $R_{\mbox{\sz ext}}$ respectively, for the two networks depicted in Fig.~\ref{f7}. Modules are sorted according to medians in increasing order. We label the horizontal axis using names for the modules assigned according to the geographical location of at least the 75\% of their nodes. We highlight whiskers and outliers in both networks. Only those modules whose structure is significant (more than 10 nodes) are represented in the plot.}
\label{f8}
\end{figure}
\end{landscape}

The structure of modules is scrutinized in Fig.~\ref{f8}, where we depict the box-and-whisker plots of the internal contributions $R_{\mbox{\sz int}}$ and external contributions $R_{\mbox{\sz ext}}$. The results show the heterogeneity of each module of the partition. Remarkably, the method reveals outliers distinguished by their capability to support the internal structure of modules and also to cross-connect them. In Fig.~\ref{f8}a (top), we observe that USA and W Europe modules have medians greater than the percentiles-75 of the rest of modules. This fact is pointing out the extreme internal cohesion of both sites. We also observe that the lowest value in $R_{\mbox{\sz int}}$ median corresponds to Alaska, however Anchorage leads the internal cohesion orders of magnitude beyond the core. In Fig.~\ref{f8}a (bottom) Canada, W Europe an C America provide the highest profile of boundary connectivity. Nevertheless, the role played by USA is still very significant because of its high percentiles and outliers. On the other side, Africa, Russia and China are less connected to the world than the rest of modules. For the AS--P2P the box-and-whisker plots in $R_{\mbox{\sz int}}$ Fig.~\ref{f8}b (top) inform about a slight dominance of 3 modules E Europe, W Europe and the module containing USA and Japan. Here E Europe does not correspond to the political area but to a tag we use to represent a geographical area that is more oriental than the western, denoted as W Europe. In $R_{\mbox{\sz ext}}$ Fig.~\ref{f8}b (bottom) the similarity in range and medians reveals the homogeneity of the mesoscale of this network. Significantly, some highlighted ASs in the plot do not belong geographically to the assigned tag, although the main proportion of nodes in that module do (see E Europe, W Europe and Russia).

\begin{figure}[!tpb]
  \begin{center}
  \begin{tabular}{l}
    {\bf a)} \hspace{25pt} Airports
    \\
    \mbox{\includegraphics*[angle=-90,width=.70\textwidth]{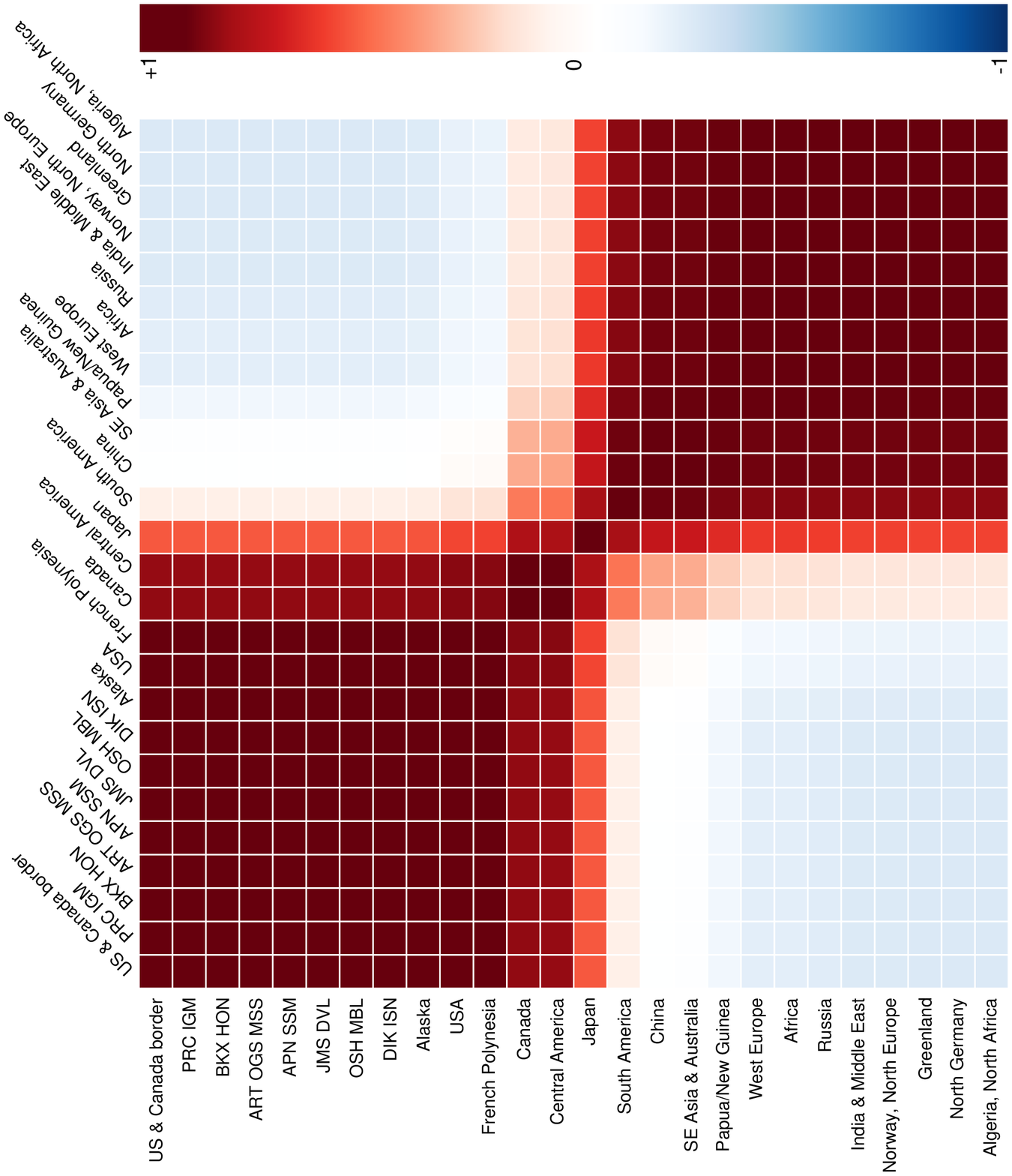}}
    \\ \\
    {\bf b)} \hspace{25pt} AS--P2P
    \\
    \mbox{\includegraphics*[angle=-90,width=.70\textwidth]{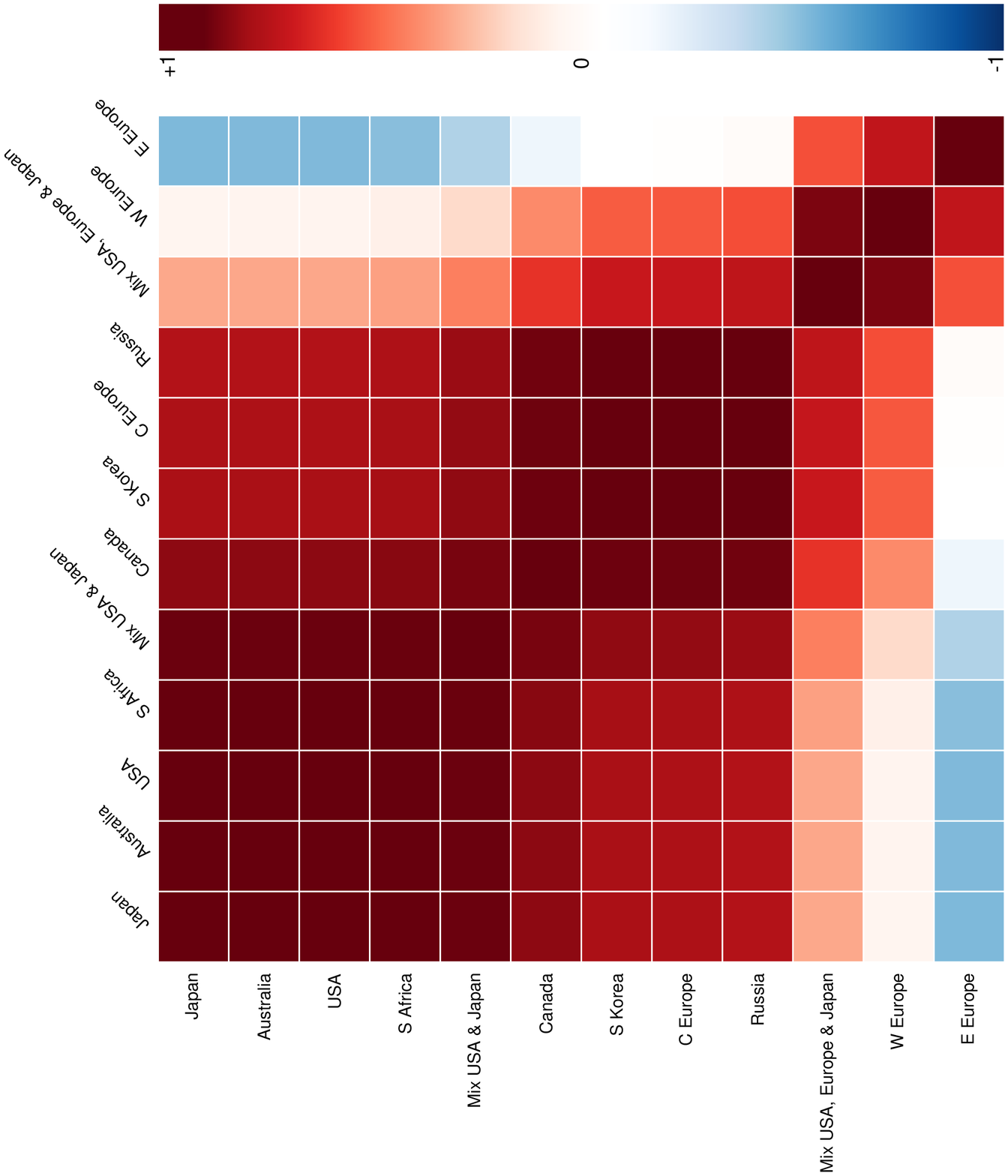}}
  \end{tabular}
  \end{center}
\caption{Overlap matrices between the modules composing the topological mesoscale of the networks plotted in Fig.~\ref{f7}. Each matrix corresponds to the normalized scalar product of the individual modular projections (see text for details). Modules are sorted by decreasing order of modular projection's angle in the plane $\mathcal{U}_2$.}
\label{f9}
\end{figure}

Finally, we plot the interrelations between modules in Fig.~\ref{f9} by computing the scalar product of their respective modular projections. The labels of the matrix are chosen in decreasing order of modular projection's angle $\theta_{\bm{\st\tilde{m}}_{\alpha}}$. For the airports network (Fig.~\ref{f9}a) we observe a clearly polarized structure in two main blocks, with a more diffuse central part overlapping both (corresponding to the communities mainly composed by nodes in Canada, Central America, Japan and South America). Japan is especially interesting for it maintains no preference in overlapping with any specific module in the network.  In the AS--P2P network (Fig.~\ref{f9}b) we observe four groups, where neighbors in the analysis are in accordance with geographical neighbors. We remark that geographical information is not included in any part of the analysis, it simply emerges from the projection of the contribution matrix. The geographical correlation in the AS--P2P network could surprise given that communities of use in P2P networks are related to contents or topics, however many AS have to pay to other ASs to provide the connection between peers and then geopolitical constraints are revealed.

\section{Conclusions}

Summarizing, we have reformulated the analysis of the modular structure first, defining the object
that contains all this information, and second we apply Singular Value
Decomposition (SVD) on this object. Dimensional reduction follows in a natural
way from the properties of the truncation of SVD, in particular we concentrate
on the truncation of rank 2, with the idea of having a map of the modular
structure amenable for analysis to any scientist. The approach is very simple
and can be understood using basic algebra notions. The computational implementation is also affordable
given the multiple software packages that include an automatic SVD (R and
Matlab among others). The result is a formalism to study the skeleton of
networks at the modular level.
The most important problem we have faced in the current research was the
interpretation of the outcome in terms of the original data. We have made a
breakthrough on this interpretation by focusing our attention in the particular
resulting geometry of the projected contribution of nodes. We also present a
statistical analysis of the resulting map using Box-and-Whisker plots based on
percentiles, more appropriate than the use of z-scores that must assume a
Gaussian distribution of values. Finally, we find the map of interrelations of the
modular skeleton.

The method proposed might be very useful for scholars in different disciplines that want access to an easy and tractable map of the empirical complex network data according to a biological, functional or topological partitions. We devise that the analysis of this map will be very helpful to anticipate the scope of dynamic emergent phenomena that depends on the structure and relations between modules. Spreading of viruses or synchronization processes are natural candidates to be analyzed considering the organization of the map. Moreover, we devise that the method can be used to graph bipartitioning
by adaptively changing nodes between two modules while maximizing the angle in the $R-\theta$ plane between them. Further studies of the similarities between nodes' contribution projections can also help to classify networks according to the role profiles of nodes \cite{rogernatphys} and/or modules.

\section*{Acknowledgements}
We acknowledge A. D{\' i}az-Guilera, R. Guimer\`a and. C. Zhou for useful discussions, also the group of Prof. L.A.N. Amaral for sharing the air transportation network data. This work was supported by Spanish Ministry of Science and Technology FIS2009-13730-C02-02 and the Generalitat de Catalunya SGR-00838-2009. A.A. acknowledges support by the Director, Office of Science, Computational and Technology Research, U.S. Department of Energy under Contract No. DE-AC02-05CH11231.G.Z.-L. was supported by the Deutsche Forschungsgemeinschaft, research group FOR 868 (contract no KU 837/23-1) and the BioSim network of excellence, contract numbers LSHB-CT-2004-005137 and -65533.

\appendix

\section{Properties of TSVD}

Let us assume that we preserve only the $r$ largest singular values and neglect the remaining substituting their value by zero, then the reduced matrix $\bm{C_r}  = \bm{U} \bm{\Sigma_r} \bm{V}^{\dag}$ has several mathematical properties worth to mention: first, it minimizes the Frobenius norm ($\|\bm{A}\|_{F}=\sqrt{\mbox{trace}(\bm{A}\bm{A}^\dag)}$) of the difference $\|\bm{C}-\bm{C_r}\|_{F}$, that means that among all possible matrices of rank $r$, $\bm{C_r}$ is the best approximation in a least squares sense; second, $\bm{C_r}$  is also the best approximation in the sense of statistics, it maintains the most significant information portion of the original matrix \cite{chu05}. The left and right singular vectors (from matrices $\bm{U}$ and $\bm{V}$ respectively) capture invariant distributions of values of the contribution of nodes to the different modules. In particular the larger the singular value the more information represented by their corresponding left and right singular vectors. We have used the LAPACK-based implementation of SVD in MATLAB. We warn that some numerical implementations of SVD suffer from a sign indeterminacy, in particular the one provided by MATLAB is such that the first singular vectors from an all-positive matrix always have all-negative elements, whose sign obviously should be switched to positive \cite{kolda}.

\section{Projection using TSVD of rank 2}

In the case of a rank $r=2$ approximation, the unicity of the two-ranked decomposition is ensured \cite{svd} if the ordered singular values $\sigma_i$ of the matrix $\bm{\Sigma}$, satisfy $\sigma_1>\sigma_2>\sigma_3$. This dimensional reduction is particularly interesting to depict results in a two-dimensional plot for visualization purposes. In the new space there are two different sets of singular vectors: the left singular vectors (columns of matrix $\bm{U}$), and the right singular vectors (rows of matrix $\bm{V}^{\dag}$). Given that we truncate at $r=2$, we fix our analysis on the two first columns of $\bm{U}$, we call this the projection space $\mathcal{U}_2$. The coordinates $\bm{\tilde{n}}_i$ of the projection of the contributions $\bm{n}_i$ of node $i$ are computed as follows:
\begin{equation}
  \bm{\tilde{n}}_i=\bm{\Sigma_2}^{-1} \bm{V}^{\dag} \bm{n}_i
\end{equation}
Here $\bm{\Sigma_2}^{-1}$ denotes the pseudo-inverse of the diagonal rectangular matrix $\bm{\Sigma_2}$ (singular values matrix truncated in 2 rows), simply obtained by inverting the values of the diagonal elements. It is possible to assess the loss of information of this projection compared to the initial data by computing the relative difference between the Frobenius norms:
\begin{equation}
  E_r =
    \frac{ \|\bm{C}\|_{F} - \|\bm{C_r}\|_{F} }{ \|\bm{C}\|_{F} } =
    \frac{\ds \sum_{\alpha=1}^{M}\sigma_{\alpha}^{2}-\sum_{\alpha=1}^{r}\sigma_{\alpha}^{2}}
         {\ds \sum_{\alpha=1}^{M}\sigma_{\alpha}^{2}}
\end{equation}

\section{Geometrical properties of the projection of $\bm{C}$}

The intramodular projection $\bm{\tilde{e}}_{\alpha}$ corresponding to module $\alpha$, is defined as the projection of the cartesian unit vector
$\bm{e}_{\alpha}=(0,\ldots,0,1,0,\ldots,0)$ (the $\alpha$-th component is 1, the rest are zero),
i.e.
\begin{equation}
  \bm{\tilde{e}}_{\alpha}= \bm{\Sigma_{2}}^{-1} \bm{V}^{\dag} \bm{e}_{\alpha}
\end{equation}
Any node in the original contribution matrix can be represented as
\begin{equation}
  \bm{n}_i=\sum_{\alpha=1}^{M} C_{i\alpha}\bm{e}_{\alpha}
\end{equation}
Its projection gives the node contribution projection
\begin{equation}
  \bm{\tilde{n}}_i=\sum_{\alpha=1}^{M} C_{i\alpha} (\bm{\Sigma_{2}}^{-1} \bm{V}^{\dag} \bm{e}_{\alpha}) = \sum_{\alpha=1}^{M} C_{i\alpha} \bm{\tilde{e}}_{\alpha}
\end{equation}
a linear combination of intramodular projections.
 In particular, a node $i$ whose contribution is totally internal to a module $\alpha$ is projected as $\ds\bm{\tilde{n}}_i=k_i \bm{\tilde{e}}_{\alpha}$, where $k_i$ is the node degree. The modular projections $\bm{\tilde{m}}_{\alpha}$ are computed as the vector sum of all the projections of nodes contributions, for those nodes belonging to module $\alpha$, i.e.
\begin{equation}
\bm{\tilde{m}}_{\alpha}=\sum_{i=1}^{N} S_{i\alpha}\bm{\tilde{n}}_i
\end{equation}

\section{Effect of noise on $\bm{C}$}
\begin{figure}[!t]
    \mbox{\includegraphics*[width=.95\textwidth]{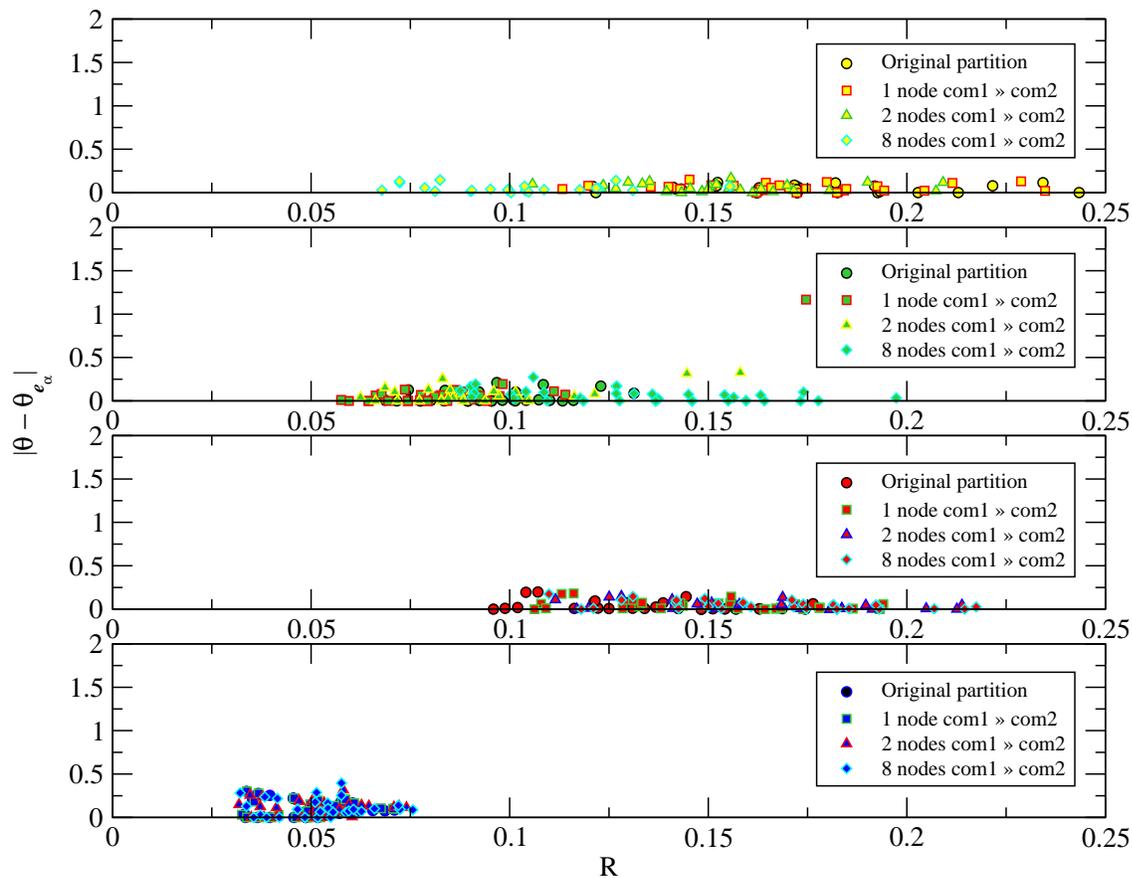}}
\caption{Robustness of the method to noise in the partition. We show the separation from the intramodular directions of modules 1 to 4 (top to down) of all nodes, in particular we track the deviation of the nodes when some of them have been assigned to the incorrect module. The nodes that have been moved are those that deviate more from the intramodular projection of module 2.}
\label{noises}
\end{figure}

The method presented is pretty robust to perturbations in the partition or, equivalently, in the contribution matrix $\bm{C}$.
To support the claim we make the following experiment: using the benchmark network proposed by Newman and Girvan \cite{gn}, see section 3, with 128 nodes, $z_{in}=15$ and $z_{out}=1$, we perform
slight changes in the predefined partition, by moving nodes from module 1 to module 2. First we move only one node, then two nodes, and finally 8 nodes.
This changes matrix $\bm{C}$, which must in turn affect TSVD output. Fig.~\ref{noises} contains the nodes' projection as the mentioned movements take place (squares, triangles and diamonds respectively). Consistently, module 1's nodes projections progressively decrease in $R$. Module 2 balances this fact, it retains the weight leaving from module 1. Sensitivity to inter-modular connections is also evidenced: when a single new node appears in module 2 (Fig.~\ref{noises}, squares), $\phi_{i}$ has an outstanding value if compared to the rest; this is also evident when two nodes enter group 2 (Fig.~\ref{noises}, triangles). When moving 8 nodes, the effect is less drastic for the deviations in $\theta$ and more drastic in $R$. Unsurprisingly, modules 3 and 4 remain mostly unchanged, the interplay between modules 1 and 2 (nodes leaving from one group towards the other) does not drastically affect their internal characteristics, nor their importance in the whole structure.


\section*{References}
\begin{thebibliography}{99}

\bibitem{gn}
Girvan M and Newman M E J 2002
Community structure in social and biological networks
{\em Proc. Natl. Acad. Sci. USA} {\bf 99} 7821

\bibitem{vespi}
Vespignani A 2003
Evolution thinks modular
{\em Nature Genetics} {\bf 35} 118

\bibitem{ng}
Newman M E J and Girvan 2004
Finding and evaluating community structure in networks
{\em Phys. Rev. E} {\bf 69} 026113

\bibitem{palla}
Palla G, Der{\'e}nyi I, Farkas I and Vicsek T 2005
Uncovering the overlapping community structure of complex networks in nature and society
{\em Nature} {\bf 435} 814

\bibitem{ddda}
Danon L, D\'iaz-Guilera A, Duch J and Arenas A 2005
Comparing community structure identification
{\em J. Stat. Mech.} P09008

\bibitem{physrep}
Arenas A, D{\' i}az-Guilera A, Kurths J, Moreno Y and Zhou C 2008
Synchronization in complex networks
{\em Physics Reports} {\bf 469} 93

\bibitem{rogernat}
Guimer\`a R and Amaral L A N 2005
Functional cartography of metabolic networks
{\em Nature} {\bf 433} 895

\bibitem{rogerair}
Guimer\`a R, Mossa S, Turtschi A , Amaral L A N 2005
The worldwide air transportation network: anomalous centrality, community structure, and cities' global roles
{\em Proc. Natl. Acad. Sci. USA}  {\bf 102} 7794




\bibitem{svd}
Golub G  H and Van Loan C F 1996
{\em Matrix Computations} 3rd ed
(Baltimore: Johns Hopkins University Press)

\bibitem{berry}
Berry M W, Dumais S T and O'Brien G W 1995
Using Linear Algebra for Intelligent Information Retrieval
{\em SIAM Review} {\bf 37} 573

\bibitem{landauer}
Landauer T and Dumais S T 1997
A solution to Plato's problem: The Latent Semantic Analysis theory of acquisition, induction, and representation of knowledge
{\em Psychological Review} {\bf 104} 211

\bibitem{chu05}
Chu M T 2005 and Golub G H
{\em Inverse eigenvalue problems: theory, algorithms, and applications}
(Oxford: Oxford University Press) pp˜279-286

\bibitem{genes1}
Alter O, Brown P O and Botstein D 2000
Singular value decomposition for genome-wide expression data processing and modeling
{\em Proc. Natl. Acad. Sci. USA} {\bf 97} 10101

\bibitem{genes2}
Langfelder P and Horvath S 2007
Eigengene networks for studying the relationships between co-expression modules
{\em BMC Systems Biology} {\bf 1} 54

\bibitem{rosenblum}
Rosenblum  M G,  Pikovsky A S, Kurths J, Sch\"afer C and Tass P 2001
Phase Synchronization: From Theory to Data Analysis,
{\em Handbook of Biological Physics, Neuro-informatics and Neural Modeling}  {\bf 279}

\bibitem{physicad}
Arenas A,  D{\' i}az-Guilera A and P\'erez-Vicente C 2006
Synchronization processes in complex networks
{\em Physica D} {\bf 224} 27

\bibitem{zachary}
Zachary W W 1977
An information flow model for conflict and fission in small groups
{\em Journal of Anthropological Research} {\bf 33} 452

\bibitem{EO}
Duch J and Arenas A 2005
Community identification using Extremal Optimization
{\em Phys. Rev. E} {\bf 72} 027104

\bibitem{meso}
Arenas A, Fern{\'a}ndez A and G{\'o}mez S 2008
Multiple resolution of the modular structure of complex networks
{\em New Journal of Physics} {\bf 10} 05039

\bibitem{caida}
Dimitropoulos X, Krioukov D, Riley G Y and Claffy K C 2006
Revealing the Autonomous System Taxonomy: The Machine Learning Approach
{\em Passive and Active Measurements Workshop (PAM)}

\bibitem{caida2}
Dimitropoulos X, Krioukov D, Fomenkov M, Huffaker B, Hyun Y, Claffy K C and Riley G 2007
AS Relationships: Inference and Validation
{\em ACM SIGCOMM Comp. Comm. Rev.} {\bf 37} 29

\bibitem{rogernatphys}
Guimer{\`a} R, Sales-Pardo M and Amaral L A N 2007
Classes of complex networks defined by role-to-role connectivity profiles
{\em Nature Physics} {\bf 3} 63

\bibitem{kolda}
Bro R, Acar E and Kolda T G 2008
Resolving the sign ambiguity in the singular value decomposition
{\em Journal of Chemometrics} {\bf 22} 135

\end{thebibliography}
\end{document}